# RESCATTERING EFFECTS IN PROTON INTERACTION WITH LIGHT NEUTRON RICH NUCLEI

E. T. Ibraeva,[1,*] A. V. Dzhazairov-Kahramanov,[1,2,†] O. Imambekov[1]

[1] *Institute of nuclear physics RK, 050032, str. Ibragimova 1, Almaty, Kazakhstan*
[2] *V. G. Fessenkov Astrophysical Institute "NCSRT" NSA RK, 050020, Observatory 23, Kamenskoe plato, Almaty, Kazakhstan*
[*] *ibraeva.elena@gmail.com*
[†] *albert-j@yandex.ru*

**Abstract:** Within the framework of the Glauber diffraction theory, the differential cross sections of the elastic $p^6$He, $p^8$Li, and $p^9$Li scattering were calculated at intermediate energies from 70 to 1000 MeV/nucleon. The use of realistic three-body wave functions $\alpha-n-n$ ($^6$He), $\alpha-t-n$ ($^8$Li), and $\alpha-t-2n$ ($^9$Li), obtained in the framework of modern nuclear models, and the expansion of the Glauber operator into a series of multiple scatterings in a form that is well adapted to the three-body nuclei configuration, allows the calculation of the matrix elements, by taking into account the rescattering from all structure components of the designated nuclei.

*Keywords*: Glauber multiple diffraction theory; three-body cluster model; differential cross section, multiple scattering.

## 1. Introduction

Helium and lithium isotopes offer a unique and very interesting example for accurate microscopic theoretical studies. The lightest double magic nucleus $^4$He is the core of the heavier isotope $^6$He, the last two neutrons of which form a halo related to the celebrated $^{11}$Li; $^8$Li and $^9$Li, which it is supposed, have a one- and two-neutron skin.

The structure of unstable isotopes is widely studied experimentally in inverse kinematics [1−9] because of the production of radioactive beams. The creation of a solid polarized proton target that is able to work at a sufficiently high temperature (~ 100 $K$) and in a weak magnetic field (~ 0.1 $T$) [8] offers the possibility of performing more precise experiments in the measurement of polarizing characteristics. The measurement of the vector analyzing power in the $p^6$He elastic scattering in inverse kinematics [9] has provoked new interest in studying the properties of He isotopes.

Theoretical research at energies of tens of MeV/nucleon is based on the optical model with various potentials: microscopic, phenomenological, folding, potential of chiral effective field theory. At energies of hundreds of MeV/nucleon the Glauber diffraction theory is usually used, as well as its approximations: High Energy and Optical Limit. In these models, various characteristics are calculated, including: root-mean-square (rms) radii, magnetic and quadruple moments, excitation spectra, proton and neutron densities, differential and total cross sections, and vector analyzing powers. Thus, the latest calculations of rms radii ($r_m^2, r_{pp}^2, r_{pn}^2$) for $^6$He are performed via the method of hyperspherical-harmonic expansion of the wave function (WF) by taking into account 3$N$ forces in the Hamiltonian [10], using the relativistic mean field approach (RMF) [11], in the optical model having used the Bethe-Brueckner-Hartree-Fock approach (BHF) [12]. The matter radii of $^{6-9}$Li isotopes are calculated in the tensor-optimized shell model (TOSM) [13] in no-core full configuration with the realistic nucleon-nucleon interaction JISP16 [14] and also in the optical model [12]. All of them are in reasonable agreement with experimental measurements. The correlation between the radius and the two-neutron separation energy of $^{6}$He was found in [10], and the importance of taking into account the 3$N$ forces was also shown.

The excitation energy spectra of He and Li isotopes were calculated in TOSM with various interaction potentials [13]. It was found that NN potential $AV8'$ describes the energy spectra better than $MN$. The role of tensor forces in the formation of the energy spectra and in the configuration of these nuclei has been established. The spectrum of low excitation energies for the lithium isotopes $^{6,7,8}$Li was calculated in [14] in complete configuration space without core with realistic NN potential JISP16.

All the known density distributions of $^{6}$He include: semi-empirical description of weakly bound nuclei [15, 16], the cluster-orbital shell model (COSMA) [17], the shell model with and without halo component [18], phenomenological [16], and those theoretically obtained for various effective interactions NLSH, NL3, NL2, and FSU Gold [11]. They are establishing the extended neutron tail, which reflects the halo structure of the neutron distribution. Halo neutrons can form spatially dineutron and cigar-like configurations, and as shown in [19−21], the prevalent configuration is dineutron.

Phenomenological neutron density distributions of $^{6,7,8,9,11}$Li isotopes show that only $^{11}$Li has the extended neutron halo structure. $^{8,9}$Li nuclei have a neutron skin [12]. The detailed picture of the neutron and proton density distribution of $^{8}$Li (together with $^{6,7}$Li) is presented in [14]. The calculation of $^{8}$Li density of various multipoles ($\rho^{(0)}$, $\rho^{(2)}$, $\rho^{(4)}$) has been carried out for the ground ($2^+$) and the excited ($4^+$) states. The forms of these densities ("prolate" for ground state and "oblate" for excited state) were shown, and the conclusion was drawn that the ground state proton monopole density has a slightly higher central value than the neutron ground state monopole density, but that it falls off much more rapidly with distance compared with the neutron density. In contrast, the ground state neutron quadrupole density is larger (than the proton) at the central band and at large distances. These neutron and proton density distributions (together with the rms radii) "provide an excellent framework for visualization of nuclear shape distortions and clustering effects" [14].

The dynamic characteristics DCS and $A_y$ of the $p$He and $p$Li scattering have been calculated in many papers. The analysis of the characteristics of the $p$He scattering was done in [9] in three models: cluster folding (supposing that $^{6}$He has non-cluster $2p+4n$-structure), nucleon folding (supposing that $^{6}$He has $\alpha-n-n$ cluster structure), and in the full microscopic model with non-local optic potential with three sets of one-particle WFs. The calculations with the phenomenological optical potential and with $\alpha-n-n$ cluster folding potential more adequately describe the experimental data.

The DCSs for the $p^{6,8}$He elastic scattering have been calculated using the relativistic mean field approach (RMF) at $E = 71$ MeV/nucleon with theoretically determined densities of $^{6,8}$He for effective interactions NLSH, NL3, NL2, and FSU Gold [11]. The difference between the predictions of the various densities at the forward angles is very small. There are differences at large angles and in the region of the minimum DCS, where densities NL2 and FSU Gold describe the experimental data better. The DCS and $A_y$ of the $p^{4,6,8}$He elastic scattering were obtained at the same energy in [22] using the eikonal approximation, the single scattering Glauber approximation and the folding model. The WFs of He isotopes were calculated by the Monte Carlo variational method. The results of the analysis show that for all nuclei, the best agreement with the experimental data is attained in the first order of the Glauber approximation, and that the folding model better reproduces $A_y$, while the calculated DCSs tend to be higher than the experimental. The Pauli blocking effect was studied; it partly suppresses the $p$He interaction that decreases DCS and improves the agreement with the experimental data. The same paper predicts the characteristics of the $p^{6,8}$He scattering at $E = 300$ MeV with a view to future experiments.

The calculations of DCS and $A_y$ of the $p^6$He scattering were carried out in [23] using the framework of the optical model. The nonlocal optical potential in the first order of the Watson multiple-scattering expansion was deduced in the work, which allows the separation of the contributions of the protons and neutrons to the structure, and naturally takes into account the contributions of the $\alpha$ core and two neutrons. COSMA code was used for the calculation of the density matrix of $^6$He. The calculation with the cluster optical potential, in which the potential for the $\alpha$ core is calculated with the NN $t$ matrix modified for the $\alpha$ particle, gives the most adequate description of $A_y$.

Systematic analysis of the $p^{4,6,8}$He and $p^{6,7,9,11}$Li scattering data at energies from 15 to 70 MeV/nucleon using the BHF method as a basis for the calculation of the optical potential were presented in [12]. Realistic internucleon NN potentials, such as Reid93, Urbana v-14, and Argonne v-18, along with several models of nucleon density distributions, are employed in generating the nucleon-nucleus optical potential.

In all the above-cited papers, the experimental DCSs reproduced rather well, while the calculated $A_y$ recently measured data of $p^6$He scattering at 71 MeV/nucleon [9] show deviation from the experimental.

In the present paper, we present the calculations of DCS of elastic scattering protons from $^6$He, $^8$Li, $^9$Li nuclei in the framework of the Glauber theory. It is necessary to know the parameters of the elementary nucleon-nucleon amplitudes and the WFs of the nuclei for calculations of the scattering matrix elements (amplitudes). Parameters of elementary nucleon-nucleon amplitudes are taken from the experiments by $pp$ and $pn$ scattering, and the WFs of nuclei are obtained from the Schrödinger equation with various potentials. We use WFs obtained in the framework of the modern three-body models $\alpha-n-n$ (for $^6$He) [24], $\alpha-t-n$ (for $^8$Li) [25], and $\alpha-t-2n$ (for $^9$Li) [26] with realistic intercluster potentials. The justification for using intercluster potentials to calculate the WFs was given in our previous papers [25–28]. The static characteristics of the nuclei (rms radii, magnetic and quadrupole moments), the binding energies in the channels, and their comparison with data were shown. This comparison has allowed us to obtain WFs that more correctly reproduce all sets of data. Having fixed the WF in such a manner, in the present paper we focus on the scattering mechanism, taking into account the full expansion series of multiple scattering in the Glauber operator $\Omega$, and we show the contribution of the high-order multiples of collision and of their interference to the DCS.

The estimates of high-order multiple scattering contributions in various formalisms were given for scattering of protons on $^{11}$Li [29], $^6$He [30], $^{11}$Be [31] nuclei. Thus, a multiple scattering expansion in the transition amplitudes for proton scattering from each projectile subsystem was used for the calculations of the DCS of the $p^{11}$Li scattering at $E = 800$ MeV/nucleon [29]. Limiting by the double scattering, the authors have shown that at small-angle scattering the DCS of single scattering is more than the sum of the single and double one. The elastic scattering of the halo nucleus $^6$He on a proton target at 717 MeV/nucleon is investigated within three different multiple-scattering formulations of the total transition amplitude [30]. A comparison of the DCSs that have been determined in the Glauber approximation, in the factorized impulse approximation (FIA), and in the fixed scatterer approximation (FSA) was done. All these calculations in the single-scattering approximation reproduced very well the data up to 0.05 (GeV/$c$)$^2$. However, at higher-momentum transfers the single-scattering cross section significantly overestimates the data, whereas the cross section including all higher-order contributions approaches the data well.

The calculations of the $p^{11}$Be elastic scattering at $E = 100$, 150, and 200 MeV/nucleon were presented in the Faddeev formalism of multiple scattering, and a comparison with the Watson statement (the expansion full transfer amplitude in multiple scattering series) and with the Glauber approximation was made in [31]. The importance of using the realistic NN interaction in any considerable calculations of scattering was emphasized. It was shown that the DCS in the single scattering approach at small angles is overestimated in comparison with estimates taking into account full expansion.

All these referenced works have drawn the conclusion that "dynamical higher-order contributions are very significant at this high-momentum transfer and need to be included" [31].

## 2. Brief formalism of the Glauber theory

According to the Glauber multiple-scattering theory, the proton elastic scattering amplitude on the nucleus with mass number $A$ can be written as the integral over the impact parameter plane $\boldsymbol{\rho}_\perp$ [32]:

$$M_{if}(\mathbf{q}_\perp) = \sum_{M_J M'_J} \frac{ik}{2\pi} \int d\boldsymbol{\rho}_\perp d\mathbf{R}_A \exp(i\mathbf{q}_\perp \boldsymbol{\rho}_\perp) \delta(\mathbf{R}_A) \left\langle \Psi_i^{JM_J} \left| \Omega \right| \Psi_f^{JM'_J} \right\rangle, \quad (1)$$

where subscript $\perp$ denotes the two-dimensional vectors (lying in a plane xy that is perpendicular to the direction of incident particles, i.e., axis z); $\Psi_i^{JM_J}, \Psi_f^{JM'_J}$ – WFs of the initial and the final states of the nucleus, in the case of elastic scattering $\Psi_i^{JM_J} = \Psi_f^{JM'_J}$; $\Omega$ – multiple-scattering operator; $\mathbf{R}_A = \frac{1}{A}\sum_{n=1}^{A}\mathbf{r}_n$ is the coordinate of the center of mass (cm) of the nucleus, $\mathbf{k}, \mathbf{k}'$ are incoming and outgoing momenta of particles in the cm system; $\mathbf{q}_\perp = \mathbf{k} - \mathbf{k}'$ is the momentum transfer.

The calculation of the matrix elements (1) for $^6$He in the $\alpha$-$n$-$n$-model was described in [28]. We will give the brief derivation of the matrix element (1) for the scattering of protons on $^9$Li, presented in the $\alpha$–$t$–$2n$ model. The conclusion for the $\alpha$–$t$–$n$-model of $^8$Li will be analogous with changing $2n \to n$.

The wave function of $^9$Li in the $\alpha$–$t$–$2n$ model is written as a product of separate clusters WFs: $\Psi_\alpha(\mathbf{R}_\alpha)$ – $\alpha$-particle, $\Psi_t(\mathbf{r}_1\mathbf{r}_2\mathbf{r}_3)$ – triton ($t$), $\varphi_{2n}(\mathbf{r}_4\mathbf{r}_5)$ – dineutron ($2n$) and the function of relative motion $\Psi_{\lambda lLS}^{JM_J}(\mathbf{r},\mathbf{R})$:

$$\Psi_{i,f}^{JM_J} = \Psi_\alpha(\mathbf{R}_\alpha)\Psi_t(\mathbf{r}_1,\mathbf{r}_2,\mathbf{r}_3)\varphi_{2n}(\mathbf{r}_4,\mathbf{r}_5)\sum_{\lambda lLS}\Psi_{\lambda lLS}^{JM_J}(\mathbf{r},\mathbf{R}). \quad (2)$$

Here, the $\mathbf{r}$ coordinate describes the relative $\alpha$–$t$ motion of the clusters; the orbital moment $\lambda$ with projection $\mu$ is conjugated to it, $\mathbf{R}$ describes the relative motion between the cm of the $\alpha$–$t$ system and the diniuteron; the orbital moment $l$ with projection $m$ is conjugated to it. Let us write that transition from the single-particle coordinates of nucleons forming triton

$\{\mathbf{r}_1, \mathbf{r}_2, \mathbf{r}_3\}$, dineutron $\{\mathbf{r}_4, \mathbf{r}_5\}$, and the $\alpha$-particle $\{\mathbf{R}_\alpha\}$ (which we consider as a structureless particle), to relative Jacobi coordinates of triton $\{\mathbf{a},\mathbf{b}\}$, dineutron $\{\mathbf{d}\}$ and cm coordinates α-particle $\{\mathbf{R}_\alpha\}$, triton $\{\mathbf{R}_t\}$, and dineutron $\{\mathbf{R}_{2n}\}$

$$\mathbf{r}_1 = \mathbf{R}_t + \frac{1}{3}\mathbf{b} + \frac{1}{2}\mathbf{a}, \; \mathbf{r}_2 = \mathbf{R}_t + \frac{1}{3}\mathbf{b} - \frac{1}{2}\mathbf{a}, \; \mathbf{r}_3 = \mathbf{R}_t - \frac{2}{3}\mathbf{b}, \; \mathbf{r}_4 = \mathbf{R}_{2n} + \frac{1}{2}\mathbf{d}, \; \mathbf{r}_5 = \mathbf{R}_{2n} - \frac{1}{2}\mathbf{d}$$
$$\mathbf{R}_9 = \frac{1}{9}\sum_{i=1}^{9}\mathbf{r}_i = \frac{1}{3}\mathbf{R}_t + \frac{2}{9}\mathbf{R}_{2n} + \frac{4}{9}\mathbf{R}_\alpha. \tag{3}$$

On condition that $\mathbf{R}_9 = 0$ we will obtain

$$\mathbf{R}_t = \frac{2}{9}\mathbf{R} - \frac{4}{7}\mathbf{r}, \; \mathbf{R}_\alpha = \frac{2}{9}\mathbf{R} + \frac{3}{7}\mathbf{r}, \; \mathbf{R}_{2n} = -\frac{7}{9}\mathbf{R}. \tag{4}$$

The relative motion WF in Jacobi coordinates, expanded into a series by partial waves, whose radial part is presented as a series on a multivariate Gaussian basis, will be written as:

$$\Psi^{JM_J}_{\lambda lLS}(\mathbf{r},\mathbf{R}) = \sum_{M_L M_S \mu m} \langle \lambda\mu l m | L M_L \rangle \langle s_1 m_1 s_2 m_2 | S M_S \rangle \langle L M_L S M_S | J M_J \rangle \times$$
$$\times Y_{\lambda\mu}(\Omega_r) Y_{lm}(\Omega_R) \chi_{SM_S} \times r^\lambda R^l \sum_{ij} C^{\lambda l}_{ij} \exp(-\alpha_i r^2 - \beta_j R^2), \tag{5}$$

where $\langle L M_L S M_S | J M_J \rangle$ – is the Clebsh-Gordan coefficient determining the composition of the momenta scheme, $Y_{\lambda\mu}(\Omega_r), Y_{lm}(\Omega_R)$ are the spherical harmonies, $\chi_{SM_S}$ is the spin function, $C^{\lambda l}_{ij}, \alpha_i, \beta_j$ are the linear and nonlinear variation parameters of task with realistic intercluster potentials of interaction.

As shown in [25, 27], three components with compounded weight of more than 95%, give the main contribution in $\Psi^{JM_J}_{\lambda lLS}(\mathbf{r},\mathbf{R})$:

$$\Psi^{JM_J}_{\lambda lLS}(\mathbf{r},\mathbf{R}) = \Psi^{JM_J}_{2121/2}(\mathbf{r},\mathbf{R}) + \Psi^{JM_J}_{1221/2}(\mathbf{r},\mathbf{R}) + \Psi^{JM_J}_{3221/2}(\mathbf{r},\mathbf{R}). \tag{6}$$

In the Glauber theory, the operator $\Omega$ is written in the form of a multiple scattering series

$$\Omega = 1 - \prod_{\nu=1}^{A}(1 - \omega_\nu(\boldsymbol{\rho}_\perp - \boldsymbol{\rho}_{\perp\nu})) = \sum_{\nu=1}^{A}\omega_\nu - \sum_{\nu<\mu}\omega_\nu\omega_\mu + \sum_{\nu<\mu<\eta}\omega_\nu\omega_\mu\omega_\eta + ...(-1)^{A-1}\omega_1\omega_2...\omega_A, \tag{7}$$

where $\boldsymbol{\rho}_{\perp\nu}$ is the two-dimensional analogue of the three-dimensional single particles coordinates of nucleons $\mathbf{r}_\nu$.

We rewrite the operator (7) in an alternative form, knowing that the scattering occurs on α-particles and that two neutrons make up the $^9$Li nucleus:

$$\Omega = \Omega_\alpha + \Omega_t + \Omega_{2n} - \Omega_\alpha\Omega_t - \Omega_\alpha\Omega_{2n} - \Omega_t\Omega_{2n} + \Omega_\alpha\Omega_t\Omega_{2n}. \tag{8}$$

Offering that α-particle is structureless, we will write the operator $\Omega_\alpha$ by the profile function $\omega_\nu$

$$\Omega_\alpha = \omega_\nu(\boldsymbol{\rho}_\perp - \mathbf{R}_{\perp\alpha}) = \frac{1}{2\pi i k} \int d\mathbf{q}_\perp \exp(-i\mathbf{q}_\perp(\boldsymbol{\rho}_\perp - \mathbf{R}_{\perp\alpha})) f_{p\alpha}(q), \tag{9}$$

where $f_{p\alpha}(q)$ is the amplitude recorded in a standard way as:

$$f_{p\alpha}(q) = \frac{k\sigma_{p\alpha}}{4\pi}(i + \varepsilon_{p\alpha})\exp\left(-\frac{(\beta_{p\alpha}q)^2}{2}\right), \tag{10}$$

with parameters $\sigma_{pn}, \varepsilon_{pn}, \beta_{pn}$ taken from papers [33−35]. Substituting (10) in (9) and integrating by $d\mathbf{q}_\perp$, we will obtain

$$\Omega_\alpha = F_\alpha \exp(-(\boldsymbol{\rho}_\perp - \mathbf{R}_{\perp\alpha})^2 \eta_\alpha), \tag{11}$$

where

$$F_\alpha = \frac{\sigma_{p\alpha}}{4\pi\beta_{p\alpha}^2}(1 - i\varepsilon_{p\alpha}), \quad \eta_\alpha = \frac{1}{2\beta_{p\alpha}^2}. \tag{12}$$

Substituting in (11) the expression for $\mathbf{R}_\alpha$ by Jacobi coordinates (4), we will obtain

$$\Omega_\alpha = F_\alpha \exp\left\{-\left(\boldsymbol{\rho}_\perp^2 + \frac{4}{81}\mathbf{R}_\perp^2 + \frac{9}{49}\mathbf{r}_\perp^2 - \frac{4}{9}\boldsymbol{\rho}_\perp\mathbf{R}_\perp - \frac{6}{7}\boldsymbol{\rho}_\perp\mathbf{r}_\perp + \frac{12}{63}\mathbf{R}_\perp\mathbf{r}_\perp\right)\eta_\alpha\right\}. \tag{13}$$

The operator $\Omega_t$ will be written in the form of a series of single, double, and triple proton collisions with triton nucleons.

$$\Omega_t = \sum_{\nu=1}^{3}\omega_\nu - \sum_{\nu<\mu}\omega_\nu\omega_\mu + \omega_1\omega_2\omega_3. \tag{14}$$

Here, the profile functions $\omega_\nu$ are expressed similarly to (9) with the replacement of the index $\alpha \to N$. After integration over $d\mathbf{q}_\perp$ and substitution of Jacobi coordinates (4) instead of single-particles coordinates, we will obtain

$$\Omega_t = \sum_{k=1}^{7} g_k \exp(-h_{1k}\boldsymbol{\rho}_\perp^2 - h_{2k}\mathbf{R}_\perp^2 - h_{3k}\mathbf{r}_\perp^2 - h_{4k}\mathbf{b}_\perp^2 - h_{5k}\mathbf{a}_\perp^2 + h_{6k}\mathbf{R}_\perp\mathbf{r}_\perp + h_{7k}\boldsymbol{\rho}_\perp\mathbf{R}_\perp + h_{8k}\boldsymbol{\rho}_\perp\mathbf{r}_\perp + \\ h_{9k}\boldsymbol{\rho}_\perp\mathbf{b}_\perp + h_{10k}\boldsymbol{\rho}_\perp\mathbf{a}_\perp + h_{11k}\mathbf{R}_\perp\mathbf{b}_\perp + h_{12k}\mathbf{r}_\perp\mathbf{b}_\perp + h_{13k}\mathbf{R}_\perp\mathbf{a}_\perp + h_{14k}\mathbf{r}_\perp\mathbf{a}_\perp + h_{15k}\mathbf{a}_\perp\mathbf{b}_\perp), \tag{15}$$

where:

$$g_k = (F_n, F_n, F_p, -F_nF_n, -F_nF_p, -F_nF_p, F_nF_nF_p),$$

$$h_{1k} = (\eta_n, \eta_n, \eta_p, 2\eta_n, (\eta_n + \eta_p), (\eta_n + \eta_p), (2\eta_n + \eta_p)),$$

$$h_{2k} = \left(\frac{4}{81}\eta_n, \frac{4}{81}\eta_n, \frac{1}{81}\eta_p, \frac{8}{81}\eta_n, \frac{4}{81}(\eta_n + \eta_p), \frac{4}{81}(\eta_n + \eta_p), \frac{4}{81}(2\eta_n + \eta_p)\right),$$

$$h_{3k} = \left(\frac{16}{49}\eta_n, \frac{16}{49}\eta_n, \frac{16}{49}\eta_p, \frac{32}{49}\eta_n, \frac{16}{49}(\eta_n + \eta_p), \frac{16}{49}(\eta_n + \eta_p), \frac{16}{49}(2\eta_n + \eta_p)\right),$$

$$h_{4k} = \left(\frac{1}{9}\eta_n, \frac{1}{9}\eta_n, \frac{4}{39}\eta_p, \frac{2}{9}\eta_n, \left(\frac{1}{9}\eta_n + \frac{4}{9}\eta_p\right), \left(\frac{1}{9}\eta_n + \frac{4}{9}\eta_p\right), \left(\frac{2}{9}\eta_n + \frac{4}{9}\eta_p\right)\right),$$

$$h_{5k} = \left(\frac{1}{4}\eta_n, \frac{1}{4}\eta_n, 0, \frac{1}{2}\eta_n, \frac{1}{4}\eta_n, \frac{1}{4}\eta_n, \frac{1}{2}\eta_n\right),$$

$$h_{6k} = \left(\frac{16}{63}\eta_n, \frac{16}{63}\eta_n, \frac{16}{63}\eta_p, \frac{32}{63}\eta_n, \frac{16}{63}(\eta_n + \eta_p), \frac{16}{63}(\eta_n + \eta_p), \frac{16}{63}(2\eta_n + \eta_p)\right),$$

$$h_{7k} = \left(\frac{4}{9}\eta_n, \frac{4}{9}\eta_n, \frac{4}{9}\eta_p, \frac{8}{9}\eta_n, \frac{4}{9}(\eta_n + \eta_p), \frac{4}{9}(\eta_n + \eta_p), \frac{4}{9}(2\eta_n + \eta_p)\right),$$

$$h_{8k} = \left(-\frac{8}{7}\eta_n, -\frac{8}{7}\eta_n, -\frac{8}{7}\eta_p, -\frac{16}{7}\eta_n, -\frac{8}{7}(\eta_n + \eta_p), -\frac{8}{7}(\eta_n + \eta_p), -\frac{8}{7}(2\eta_n + \eta_p)\right),$$

$$h_{9k} = \left(\frac{2}{3}\eta_n, \frac{2}{3}\eta_n, -\frac{4}{3}\eta_p, \frac{4}{3}\eta_n, \left(\frac{2}{3}\eta_n - \frac{4}{3}\eta_p\right), \left(\frac{2}{3}\eta_n - \frac{4}{3}\eta_p\right), \left(\frac{4}{3}\eta_n - \frac{4}{3}\eta_p\right)\right),$$

$$h_{10k} = (\eta_n, -\eta_n, 0, 0, \eta_n, -\eta_n, 0),$$

$$h_{11k} = \left(\frac{4}{27}\eta_n, \frac{4}{27}\eta_n, \frac{8}{27}\eta_p, \frac{8}{27}\eta_n, \frac{4}{27}(-\eta_n + 2\eta_p), \frac{4}{27}(-\eta_n + 2\eta_p), \frac{8}{27}(-\eta_n + 2\eta_p)\right),$$

$$h_{12k} = \left(\frac{8}{21}\eta_n, \frac{8}{21}\eta_n, -\frac{16}{21}\eta_p, -\frac{16}{21}\eta_n, -\frac{8}{21}(-\eta_n + 2\eta_p), -\frac{8}{21}(-\eta_n + 2\eta_p), -\frac{16}{21}(-\eta_n + 2\eta_p)\right),$$

$$h_{13k} = \left(\frac{2}{9}\eta_n, \frac{2}{9}\eta_n, 0, 0, \frac{2}{9}\eta_n, \frac{2}{9}\eta_n, 0\right),$$

$$h_{14k} = \left(\frac{4}{7}\eta_n, -\frac{4}{7}\eta_n, 0, 0, \frac{4}{7}\eta_n, -\frac{4}{7}\eta_n, 0\right),$$

$$h_{15k} = \left(-\eta_n, \eta_n, 0, 0, -\frac{1}{3}\eta_n, \frac{1}{3}\eta_n, 0\right).$$

$F_n, \eta_n$ are obtained by the formulas, which are analogous to (12) with the replacement of index $\alpha$ to $N(n,p)$ and depend on the parameters of elementary amplitude $f_{pN}(q)$, taken from [36]. The summation over k in (15) denotes the summation over the scattering order: k = 1–3 are single collisions, k = 4–6 are double collisions and k = 7 are triple collisions.

Operator $\Omega_{2n}$ describes the dineutron collisions with two nucleons:

$$\Omega_{2n} = \sum_{v=4}^{5} \omega_v - \omega_4 \omega_5. \qquad (16)$$

Using the same technique as in the derivation of $\Omega_t$, integrating over $d\mathbf{q}_\perp$ and changing the single-particle coordinates to relative (4), we obtain

$$\Omega_{2n} = \sum_{m=1}^{3} u_m \exp\left\{-t_{1m}\boldsymbol{\rho}_\perp^2 - t_{2m}\mathbf{R}_\perp^2 - t_{3m}\mathbf{r}_\perp^2 + t_{4m}\boldsymbol{\rho}_\perp\mathbf{R}_\perp + t_{5m}\boldsymbol{\rho}_\perp\mathbf{d}_\perp + t_{6m}\mathbf{R}_\perp\mathbf{d}_\perp\right\}, \quad (17)$$

where $m = 1-2$ describes single dimension collisions and $m = 3$ describes the double.

$$u_m = (F_n, F_n, -F_n F_n,), \quad t_{1m} = (\eta_n, \eta_n, 2\eta_n), \quad t_{2m} = \left(\frac{49}{81}\eta_n, \frac{49}{81}\eta_n, 2\cdot\frac{49}{81}\eta_n\right),$$

$$t_{3m} = \left(\frac{7}{36}\eta_n, \frac{7}{36}\eta_n, 2\cdot\frac{7}{36}\eta_n\right), \quad t_{4m} = \left(\frac{14}{9}\eta_n, \frac{14}{9}\eta_n, 2\cdot\frac{14}{9}\eta_n\right),$$

$$t_{5m} = \left(\frac{7}{9}\eta_n, -\frac{7}{9}\eta_n, 0\right), \quad t_{6m} = \left(-\frac{7}{9}\eta_n, \frac{7}{9}\eta_n, 0\right).$$

Because the WF of $^9$Li (5) and operators (13), (15), and (17) are written in the form of expansion in Gaussian functions in relative coordinates, the further calculation of the matrix element (1) is not difficult and can be done analytically. An example of such a calculation for the $p^6$He system is given in [28].

The experimentally measured differential scattering cross section with which we compare the results obtained here, is determined by the squared absolute value of the respective matrix element:

$$\frac{d\sigma}{d\Omega} = \frac{1}{2J+1}\left|M_{if}(\mathbf{q}_\perp)\right|^2. \quad (18)$$

To estimate the contribution of scattering on the core components in the DCS of the subsystems (clusters), we substitute $\Omega$ in the form of series (8) to (18):

$$\frac{d\sigma}{d\Omega} = \frac{1}{2J+1}\left|M_{if}^{(1)}(\mathbf{q}_\perp) - M_{if}^{(2)}(\mathbf{q}_\perp) + M_{if}^{(3)}(\mathbf{q}_\perp)\right|^2, \quad (19)$$

where

$$M_{if}^{(1)}(\mathbf{q}_\perp) = \frac{ik}{2\pi}\sum_{M_J M'_J}\int d\boldsymbol{\rho}_\perp d\mathbf{R}_A \exp(i\mathbf{q}_\perp\boldsymbol{\rho}_\perp)\delta(\mathbf{R}_A)\left\{\left\langle \Psi_{\lambda lL}^{JM_J}\left|\Omega_\alpha + \Omega_t + \Omega_{2n}\right|\Psi_{\lambda lL}^{JM'_J}\right\rangle\right\}, \quad (20)$$

$$M_{if}^{(2)}(\mathbf{q}_\perp) = \frac{ik}{2\pi}\sum_{M_J M'_J}\int d\boldsymbol{\rho}_\perp d\mathbf{R}_A \exp(i\mathbf{q}_\perp\boldsymbol{\rho}_\perp)\delta(\mathbf{R}_A)\left\{\left\langle \Psi_{\lambda lL}^{JM_J}\left|\Omega_\alpha\Omega_t + \Omega_\alpha\Omega_{2n} + \Omega_t\Omega_{2n}\right|\Psi_{\lambda lL}^{JM'_J}\right\rangle\right\}, \quad (21)$$

$$M_{if}^{(3)}(\mathbf{q}_\perp) = \frac{ik}{2\pi}\sum_{M_J M'_J}\int d\boldsymbol{\rho}_\perp d\mathbf{R}_A \exp(i\mathbf{q}_\perp\boldsymbol{\rho}_\perp)\delta(\mathbf{R}_A)\left\{\left\langle \Psi_{\lambda lL}^{JM_J}\left|\Omega_\alpha\Omega_t\Omega_{2n}\right|\Psi_{\lambda lL}^{JM'_J}\right\rangle\right\}. \quad (22)$$

Here, $M_{if}^{(1)}(\mathbf{q}_\perp)$, $M_{if}^{(2)}(\mathbf{q}_\perp)$, $M_{if}^{(3)}(\mathbf{q}_\perp)$ are the partial amplitudes of single, double, and triple collisions.

## 3. Analysis of results

The DCSs of the proton elastic scattering on $^6$He, $^8$Li, $^9$Li nuclei at $E \sim 70$, 700 and 1000 MeV/nucleon are calculated according to the formulas given in the previous section.

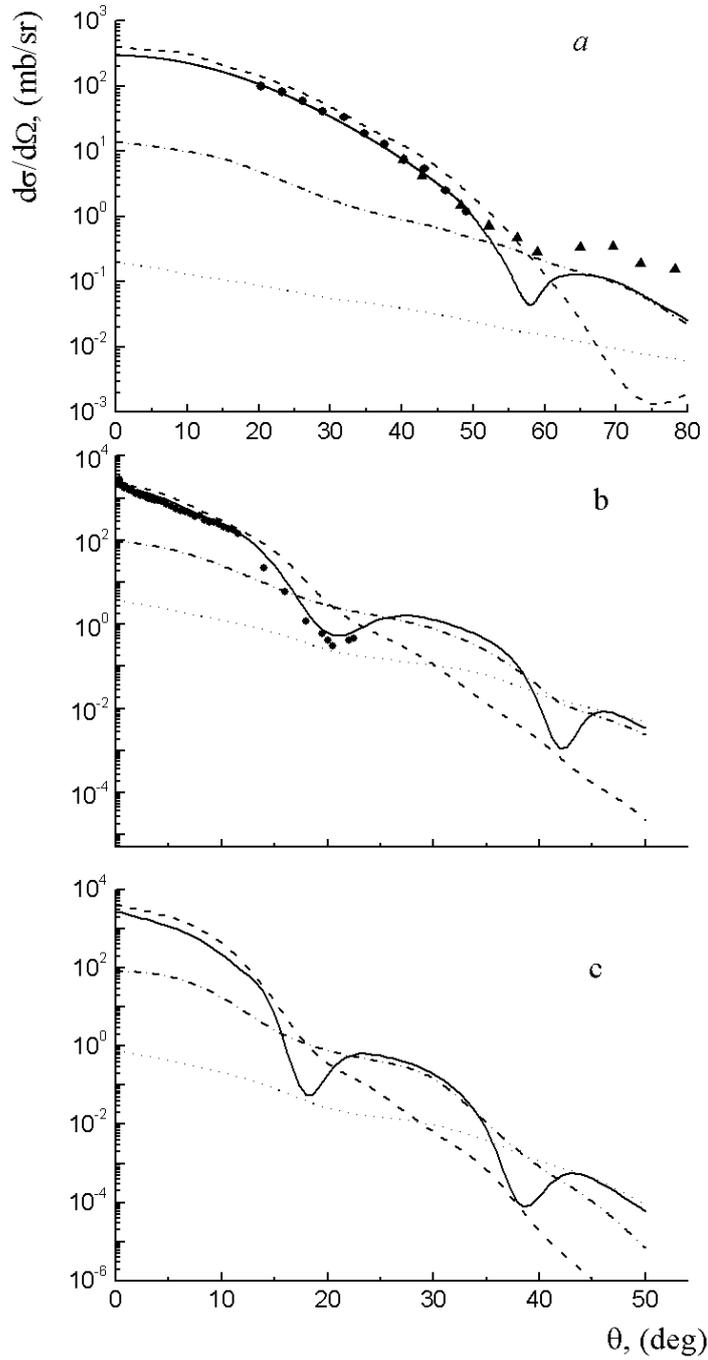

**Fig. 1.** The contributions in DCS $p^6$He-scattering single collisions (the dashed curve), double collisions (the dashed-dot curve), triple collisions (the dotted curve) and their sum (the solid curve); *a)* $E$ = 70 MeV/nucleon, the experimental data: points [3], triangles [8]; *b)* $E$ = 717 MeV/nucleon, the experimental data from [2, 4, 6]; *c)* $E$ = 1000 MeV/nucleon.

Fig. 1 shows the DCSs of the $p^6$He scattering at $E = 71$ (*a*), 717 (*b*), and 1000 (*c*) MeV/nucleon and the contribution to the cross section of the single, double, and triple rescatterings. Here, the dashed curve is the partial DCS of single collisions (contribution of the first term of expression (19) with operator $\Omega_\alpha + \Omega_n + \Omega_n$ into (20)); the dashed-dot curve is the partial DCS of double collisions (contribution of the second term of expression (19) with operator $\Omega_\alpha\Omega_n + \Omega_\alpha\Omega_n + \Omega_n\Omega_n$ into (21)); and the dotted curve is the partial DCS of triple collisions (contribution of the third term of (19) with operator $\Omega_\alpha\Omega_n\Omega_n$ into (22)). The solid curve is the aggregated contribution of all terms of expression (19), taking into account their interference. The experimental data in Fig. 1*a* are shown by points − [3] and triangles − [8], and in Fig. 1b they are from [2, 4, 6].

As seen from the figures, the single scattering gives the main contribution in the forward range of angles, but it decreases quickly and that the double and, in the sequel, triple scattering start to dominate at larger angles. The minimums, caused by the destructive interference, appear in points where the partial cross sections are crossing, because the series of multiple scattering is sign-changing (see Eqs. (7), (8), and (19)). In all cases, the summed cross section at small angles is lower than the partial cross section of single scattering; thus, the small (in this range) contribution deducts from the summed one. It can be seen in Fig. 1*a*, and 1*b* that such a cross section, decreasing in the forward range of angles, improves the agreement with the experimental data. The DCS's minimums shift to a lower range of scattering angles and the interference pattern of the cross section intensifies.

We have calculated the DCS for angles up to $\theta \sim 80°$ because of the availability of experimental data, while the calculation in the Glauber approximation makes sense only in the forward range of angles; therefore, the discrepancy between theory (solid curve) and observed experimental data is quite reasonable. Let us note that minimums in the summed DSC could be deeper, but that the first minimum is partially filled up by the contribution of the DCS of triple scattering and the second (see Fig. 1*b* and 1*c*) by the contribution of the DCS of single scattering. This implies that high-order multiples of collision give their contribution across the entire range of angles, and that their role increases with the size of the scattering angle. The similar conclusion was done in works [29−31], where the contribution of high-orders of series of multiple collisions into cross section was analyzed in the case of proton scattering on $^6$He, $^{11}$Li, $^{11}$Be nuclei in the inverse kinematics.

Fig. 2 shows the DCSs of the $p^8$Li scattering at $E = 60$ (*a*), 698 (*b*), and 1000 (*c*) MeV/nucleon taking into account contributions for collisions of different orders. The dashed, dashed-dot and dotted curves are the partial DCSs of single scattering with operator $\Omega_\alpha + \Omega_t + \Omega_n$, double scattering with operator $\Omega_\alpha\Omega_t + \Omega_\alpha\Omega_n + \Omega_t\Omega_n$, and triple scattering with operator $\Omega_\alpha\Omega_n\Omega_n$. The solid curve is the summed contribution with full operator $\Omega$. The experimental data in Fig. 2*b* are from [5].

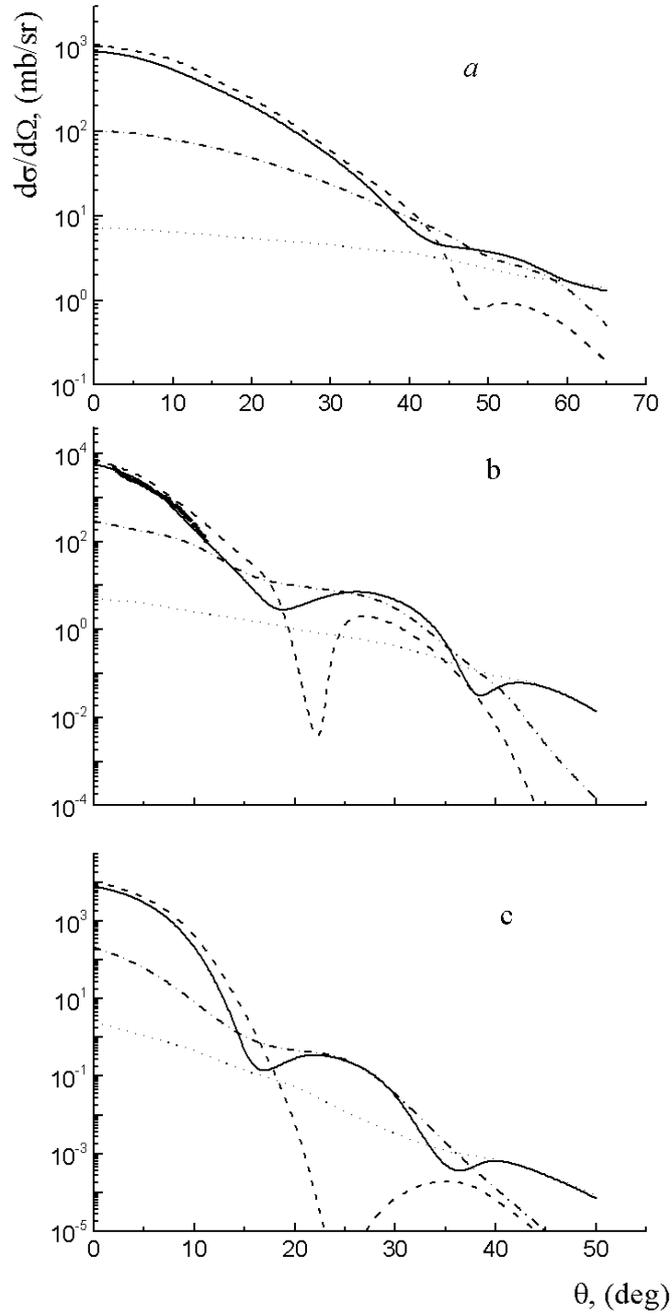

**Fig. 2.** The same that is in the Fig.1 for $p^8$Li-scattering; *a)* $E = 60$ MeV/nucleon; *b)* $E = 698$ MeV/nucleon, the experimental data from [5]; *c)* $E = 1000$ MeV/nucleon.

Fig. 3 shows the DCSs of the $p^9$Li scattering at $E = 60$ (*a*), 703 (*b*), and 1000 (*c*) MeV/nucleon taking into account contributions for collisions of different orders. The dashed, dashed-dot and dotted curves are the DCSs of single (first term of expression (19) with operator (20)), double (second term of expression (19) with operator (21)), and triple (third term of expression (19) with operator (22)) collisions. The solid curve is the summed curve, taking into account all terms of expression (19). The experimental data in Fig. 3a are from [1] and in Fig. 3b are from [5].

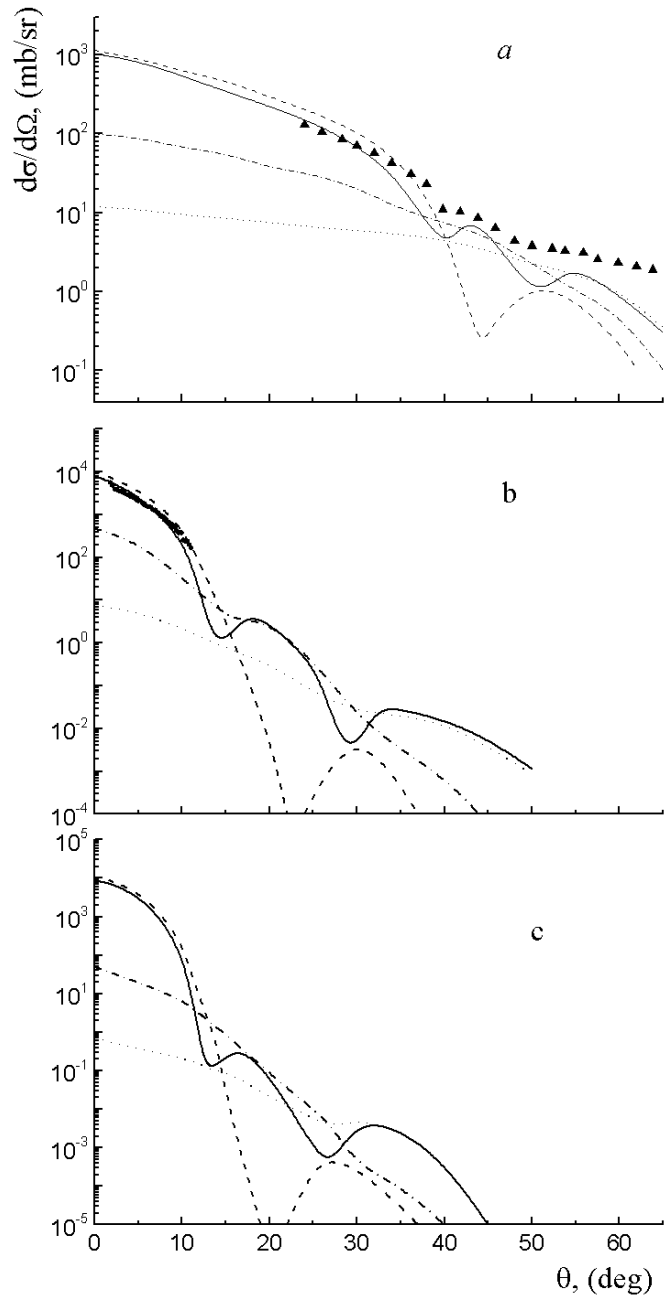

**Fig. 3.** The same that is in the Fig.1 for $p^9$Li-scattering. *a)* $E$ = 60 MeV/nucleon, the experimental data from [1]; *b)* $E$ = 703 MeV/nucleon, the experimental data from [5]; *c)* $E$ = 1000 MeV/nucleon.

The situation in Figs. 2 and 3 is analogous to that considered for the $p^6$He scattering; dominant contribution of small-angle single scattering, interference minimums in intersection points of partial cross sections of different orders, and smoothing of minimums by contributions of high orders. The available differences; two minimums in the DCS of the $p^9$Li scattering at $E = 60$ MeV/nucleon (Fig. 3a) and the deeper minimums in the partial cross sections of single scattering in Fig. 2b and 2c and Fig. 3b and 3c are caused by the structural characteristics of $^8$Li and $^9$Li nuclei in the $\alpha-t-n$ and $\alpha-t-2n$ models.

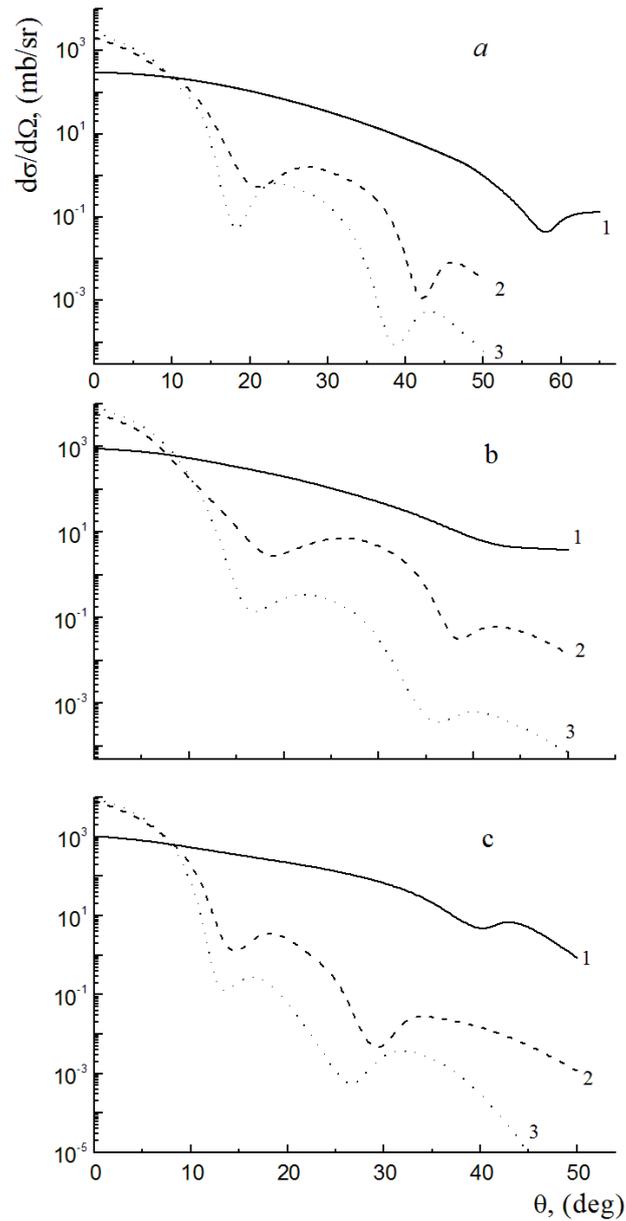

**Fig. 4.** The differential cross sections for $p^6$He (*a*), $p^8$Li (*b*), and $p^9$Li (*c*) scattering at various energies. The curves *1*, *2* and *3* are for $E = 70$ (60), 717 (689) and 1000 MeV/nucleon.

The change in DCSs with the increase of the energy of colliding particles is shown in Fig. 4: *a)* for $p^6$He, *b)* for $p^8$Li, and *c)* for $p^9$Li scattering. Curves *1*, *2* and *3* are the same curves as the solid curves (summed) in Figs. 1−3 at $E \sim 70$ (curve *1*), 700 (curve *2*), and 1000 (curve *3*) MeV/nucleon. It is seen from these figures that for all nuclei, the intensification of the diffraction pattern is observed with increasing energy: from monotonically decreasing cross section at $E \sim 70$ MeV/nucleon (curve *1*), to the cross section with two minimums and three maximums at $E \sim 1000$ MeV/nucleon (curve *3*). Such behavior of the DCSs is interpreted as the increasing energy of the particle means that it can penetrate deeper into the interior of the nucleus and can be rescattered by a greater number of nucleons.

## 4. Conclusion

We have presented the Glauber operator of multiple scattering on nucleons of nuclei in an alternative form of a series of scattering on clusters (and nucleons) that make up the nuclei. We have calculated the single, double, triple, and summed DCS of the $p^6$He, $p^8$Li, and $p^9$Li scattering at energies $E = 70$ (60), 717 (698), and 1000 MeV/nucleon and compared them with the available experimental data [1–6, 8].

It is shown that the cross section of single scattering dominates at small angles, whereas the contribution of double and triple collisions is on the scale of one and two orders less. After the first interference minimum, the contribution of the double scattering becomes dominant and after the second minimum, the triple scattering becomes dominant. Interference minimums are partially filled up by the contribution of the highest multiples of collisions, which makes the behavior of the DCS smoother, approximating it to the experimental.

The minimums in the DCS shift to the range of smaller scattering angles, the number of minimums and maximums increases in the same angular range, and the diffraction pattern becomes clearer with the increase of particle energy.

The calculation in the optical limit, when only single collisions are considered, overstates the DCS in the front angles, compared with the experimental. The contribution of double collisions reduces the cross section in this area, because it has a negative sign (see (19)), which also improves the agreement with experiment.

Thus, we can conclude that the dynamic contribution of the highest multiples of collisions must be considered at all scattering angles (momentum transfers), especially at the limit for the Glauber theory.

## Acknowledgments

This work was supported by the Grant Program of the Ministry of Education and Science of the Republic of Kazakhstan 1125/GF and 1067/GF-2.